\documentclass{jfm}
\usepackage{amsmath,amssymb}      
\usepackage{graphicx,color,subfigure,bm}
\begin{document}

\newtheorem{lemma}{Lemma}
\newtheorem{corollary}{Corollary}

\shorttitle{Vortex rings crossing a density gradient} 
\shortauthor{Y. Su \etal } 

\title{Asymmetry of motion: vortex rings crossing a density gradient}

\author
 {Yunxing Su\aff{1}, 
 Monica M. Wilhelmus\aff{1}\corresp{\email{mmwilhelmus@brown.edu}},
 Roberto Zenit\aff{1}
  }

\affiliation
{\aff{1} Center for Fluid Mechanics, Brown University, Providence RI 02912, USA}

\maketitle

\begin{abstract}

Vortex rings are critical for thrust production underwater. In the ocean, self-propelled mesozooplankton generate vortices while swimming within a weakly stratified fluid. While large-scale biogenic transport has been observed during vertical migration in the wild and lab experiments, little focus has been given to the evolution of induced vortex rings as a function of their propagation direction relative to the density gradient. In this study, the evolution of an isolated vortex ring crossing the interface of a stable two-layer system is examined as a function of its translation direction with respect to gravity. The vortex ring size and position are visualized using Planar Induced Fluorescence (PLIF) and the induced vorticity field derived from Particle Image Velocimetry (PIV) is examined. It is found that the production of baroclinic vorticity significantly affects the propagation of vortex rings crossing the density interface. As a result, any expected symmetry between vortex rings traveling from dense to light fluids and from light to dense fluids breaks down. In turn, the maximum penetration depth of the vortex ring occurs in the case in which the vortex propagates against the density gradient due to the misalignment of the pressure and density gradients. Our results have far-reaching implications for the characterization of local ecosystems in marine environments. 

\end{abstract}


\section{Introduction}\label{sec:intro}
Vortex rings are a key feature of underwater biological propulsion, particularly within the moderate Reynolds number regime. Jellyfish, for example, have consistently been observed to create vortices in their wakes during forward swimming leading to so-called biogenic transport \citep{katija2008situ,Katija2009,breitburg2010ecosystem,costello2021hydrodynamics}. While this process has been widely studied in the context of homogeneous fluids, less focus has been given to the more realistic case of vortex propagation in stratified fluids. Even though the stratification in the upper ocean is weak, changes in density are non-negligible for many species of mesozooplankton \citep{kirillin2012modeling,cullen2015subsurface,briseno2020comparing,cheneffects2022}. Here, we study the fundamental problem of vortex propagation across the interface between two miscible fluids and highlight the role of baroclinic vorticity, which originates from the misalignment of density and pressure surfaces, in impeding the penetration of vortices, depending on the direction of the ring, with important consequences for feeding and transport in ocean ecosystems.

In the fluid mechanics literature, the motion of vortex rings in density stratified fluids has been widely explored. The classical study by \citet{linden1973} experimentally analyzes the interaction of a vortex ring impinging normally on a density interface between two liquids, where a vortex ring of light fluid crosses the interface into a heavy fluid. Linden reported that the maximum penetration depth of the vortex ring was a function of the Froude number, $Fr = \rho_j U^2_j/[(\rho_0-\rho_j)gD]$, where $\rho_j$ and $U_j$ are the jet density and velocity, respectively; $\rho_0$ is the ambient fluid density and $D$ is the nozzle diameter of the vortex ring generator. Similarly, \citet{dahm1989} performed experiments with a vortex ring crossing the interface of a stable two-layer system. Experiments were conducted varying the density across the interface and producing vortex rings with different circulation values. The authors found that in the Boussinesque limit the generation of baroclinic vorticity and the subsequent evolution of the vortex ring was governed only by the product of two dimensionless parameters, $A\gamma$, where $A$ measures the density difference across the interface, and $\gamma$ measures the strength of the vortex ring. This dimensionless parameter $A\gamma$ was observed to be linearly related to the square of the Froude number, $A\gamma \sim Fr^2$. For large $A\gamma$ values, the vortex ring can barely penetrate the interface, which then acts like a solid wall. The authors also observed an inverse flow after the vortex ring crossed the interface.   

From a theoretical standpoint, \citet{saffamn1992} studied the vertical translation of a vortex ring moving against buoyancy. Consider the fluid inside the vortex core to be $\rho_1$ and the ambient fluid to be $\rho_0$, where $\rho_0 \neq \rho_1$. If the circulation around the ring is $\kappa$, then one would expect $D\kappa/Dt=0$ according to Kelvin's circulation theorem. The hydrodynamic impulse of the vortex ring would then be given by
\begin{equation}
I=\rho_1\kappa R^2,
\end{equation}
where $R$ is the vortex ring radius.
The buoyant force on the ring would be
\begin{equation}
F_b=(\rho_0-\rho_1)2\pi g Ra^2,
\end{equation}
where $a$ is the vortex core radius. If $a<<R$, it can be considered that the entrainment is negligible, so $\rho_1= constant$. {As the momentum of the vortex ring decreases due to the buoyant force, the time rate of change of momentum of the vortex ring is always negative and is balanced by the buoyant force, which should also be negative, therefore, we have
\begin{equation}
\frac{dI}{dt}= \rho_1 \kappa 2R \frac{dR}{dt} = -|\rho_0-\rho_1|2\pi g Ra^2.
\end{equation}
As a result, the evolution of the vortex ring radius is given as
\begin{equation}
R=R_0 - \frac{|\rho_0-\rho_1|}{\rho_1}\frac{\pi g}{\kappa}a^2 t. \label{eqn: vortex ring size}
\end{equation} 
The second term on the right-hand-side of Eqn.~\ref{eqn: vortex ring size} remains negative as long as there exists a density difference. While this model captures how the diameter of the vortex ring decreases as it moves through a fluid with different density, it assumes no fluid entrainment into the vortex ring. Nevertheless, the analysis can be used to lead a qualitative discussion regarding the evolution of the vortex ring size.}

\citet{marugancruz2009,marugancruz2013} studied the formation of vortex rings in a negatively buoyant environment as a function of Froude number. Specifically, a fluid was injected into a denser one using a vortex ring generator at the top of a tank. The authors found that there exists a critical Froude number below which the leading vortex ring is pushed back by the buoyant force before it develops and detaches from the injection orifice. In addition, the authors observed a thin layer of baroclinic vorticity of opposite sign to the vorticity around the vortex ring. This thin layer of vorticity was observed to move against the travel direction of the ring due to buoyancy. \cite{camassa2013numerical} experimentally and numerically studied a vortex ring settling in a two-layer configuration of miscible fluids and found that, depending on the initial conditions (e.g., vortex size, speed, and the initial distance to the interface), the the vortex ring can either be trapped in one density layer or penetrate the interface. More recently, \citet{Olsthoorn_dalziel_2015,Olsthoorn_dalziel_2017} experimentally investigated the mixing efficiency and stability effects of a vortex ring impinging a density interface by reconstructing full three-dimensional velocity field measurements. The authors also reported a time scale for the evolution of baroclinic vorticity production when a vortex ring crosses the density interface. It should be noted that \citet{scasse2005} analyzed a vortex ring crossing a density interface at an oblique angle. In this case, a three-dimensional vortex ring collapses into a nearly two-dimensional flow as a result of the stable fluid stratification.

 While most literature focuses on the important cases of turbulent mixing of stratified fluids (e.g., \cite{caulfield2021layering,smith2021turbulence}) and the evolution of vortex rings crossing a density interface (e.g., \cite{camassa2016variable}), the question remains of whether the motion of a vortex ring is symmetrical as it travels along or against a stable density gradient. Given the hundred-meter-long migrations that mesozooplankton undergo in the ocean \citep{bianchi2013intensification,bianchi2013diel}, accurate assessment of biogenic induced transport is essential to better understand the sustainability of marine ecosystems. In this study, we examine the process of a vortex ring crossing a density interface between two miscible fluids in two different directions, along and against the density gradient. The penetration distance of the vortex ring is tracked and compared between the two cases in Sect.~\ref{Sect. discription of the process}. The associated baroclinic vorticity generated at the interface is examined in Sect.~\ref{sect: penetration and circulation} and then used in further discussions on the source of asymmetry in Sect.~\ref{further discussion}.

\section{Experimental setup and methods}\label{sec:setup} 

\begin{figure}
\centering

\includegraphics[width=1\textwidth]{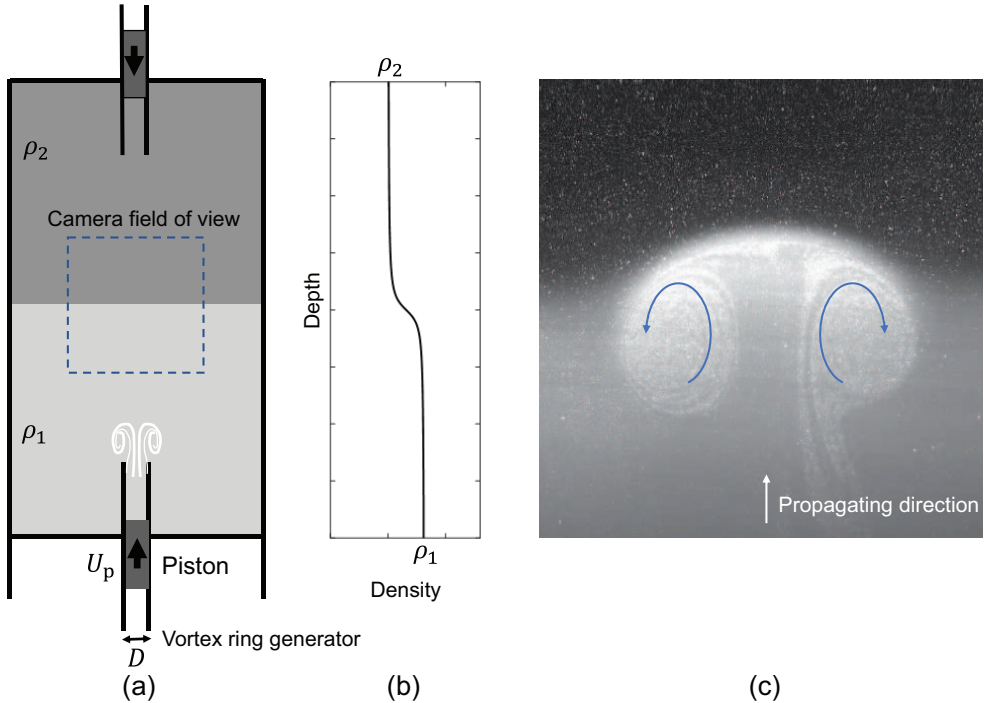}
\caption{(a) Sketch of the experimental setup consisting of an acrylic tank and two vortex generators. The tank was filled with two uniform fluid layers such that a stable stratification was achieved, i.e. $\rho_1>\rho_2$. Vortex rings that are generated from the bottom (top) of the tank propagate upward (downward) and cross the density interface. The camera field of view (boxed region) is centered at the density interface. (b) Sketch of the density variation across the depth of the tank to illustrate the density interface in (a). (c) Representative raw image of an upward propagating vortex ring crossing the density interface. The small bright dots are the tracer particles used for PIV measurements. The blue arrows indicate the recirculating direction of the flow in the vortex ring. }
\label{fig: setup}
\end{figure}

The experimental setup consists of a transparent acrylic tank (25.4$\times$25.4$\times$50.8 cm$^3$) and two identical piston-cylinder arrangements (one on top and one at the bottom) positioned at the center of the tank as shown in Fig.~\ref{fig: setup}(a). In each vortex generation system, the inner diameter of the cylinder, $D$, is 2.54 cm. The piston, which has a cylindrical cross-section and is 25.4 cm in length, moves freely within the cylinder. The motion of the piston is controlled by a hydraulic circuit, which displaces a prescribed fluid column at a given speed and distance, $L$, using a larger piston and a step motor controlled by an Arduino board. 

For all experiments, a stable two-layer density stratification was produced by slowly filling the tank midway with homogeneous fluid and subsequently adding another fluid layer with greater density. In all cases, the interface between the two fluids remained nearly stagnant during the filling process. Nonetheless, the tank was left undisturbed for about half an hour before performing experiments. To generate vortex rings, one of the pistons was set into motion with a constant speed of $U$ = 7.54 $\pm$ 0.18 cm/s. The corresponding Reynolds number was $Re=U D \rho/\mu$= 1915, considering the properties of fresh water. 
To restrict our tests to the case of individual vortex rings without a wake \citep{gharib1998}, the stroke ratio, $L/D$, for all cases presented here was 2.55.

The properties of the fluid solutions tested in this study are presented in Table \ref{table:props}. In all cases, tap water was used as the baseline, and table salt ($Na Cl$) was added to increase the density of the solution. This property was measured with a floating hydrometer with a resolution of 0.5 kg/m$^3$. The density difference encountered by the vortex ring as it crosses the interface is characterized by a normalized density contrast: $\Delta \rho^*_{21}=(\rho_2-\rho_1)/{\rho_2}$ and $\Delta\rho^*_{12}= (\rho_1-\rho_2)/\rho_1$, where $\rho_1$ and $\rho_2$ are the densities of bottom (dense) and the top (the baseline fluid, light) fluids, respectively.  As discussed in Sections 3 and 4, the value and sign of $\Delta\rho^*$ plays an important role in the motion of vortex rings (Table \ref{table:props}). 
The maximum change of viscosity between the two fluids does not surpass 3.5\% according to the data in literature \citep{qasem2021comprehensive}. This viscosity effect is, therefore, assumed to be negligible \citep{mostafa2010}.

\begin{table}
 \begin{center}
  \begin{tabular}{lcccc}
    $\rho$ (kg/m$^3$)     &   S (g/L) & $\Delta\rho^*_{12}$ & $\Delta\rho^*_{21}$  \\
    \hline
       1002   & 4.6 & 1.9$\times10^{-3}$ & -2.0$\times10^{-3}$ \\
       1004   & 7.4 & 3.9$\times10^{-3}$ & -4.0$\times10^{-3}$  \\
       1005   & 9.0 & 4.9$\times10^{-3}$ & -5.0$\times10^{-3}$\\
       1007   & 10.2 & 6.9$\times10^{-3}$ & -7.0$\times10^{-3}$ \\
       1008   & 15.7 & 7.9$\times10^{-3}$ & -8.0$\times10^{-3}$\\
       1013   & 23.6 & 1.3$\times10^{-2}$ & -1.3$\times10^{-2}$ \\
\hline
  \end{tabular}
  \caption{Properties of the liquids used in the experiments. The first column lists the density of the solutions (pure water is the top layer fluid and the baseline solution, $\rho_2$), followed by the corresponding salinity in the second column. The third and fourth columns indicate the normalized density contrasts, where $\Delta\rho^*_{12}= (\rho_1-\rho_2)/\rho_1$ and $\Delta \rho^*_{21}=(\rho_2-\rho_1)/{\rho_2}$.}{\label{table:props}}
 \end{center}
\end{table}

\subsection{Measurement techniques}

Two experimental techniques were implemented in this study. Planar Laser-Induced Fluorescence (PLIF) was used to track the location of the vortex and the density gradient, while two-dimensional Particle Image Velocimetry (PIV) was used to measure the velocity field in the mid-plane of the experimental tank. The first technique was performed by adding a small amount of fluorescent dye (Rhodamine 6G, Sigma-Aldrich) to one of the fluid layers (20 ppm). Note that since the amount of mixing during the initial instants of the interaction of the vortex ring with the interface is very small, the location of the interface was used as a proxy for the sharpest density gradient, $\nabla \rho$. 

\begin{figure}
\centering

\includegraphics[width=1\textwidth]{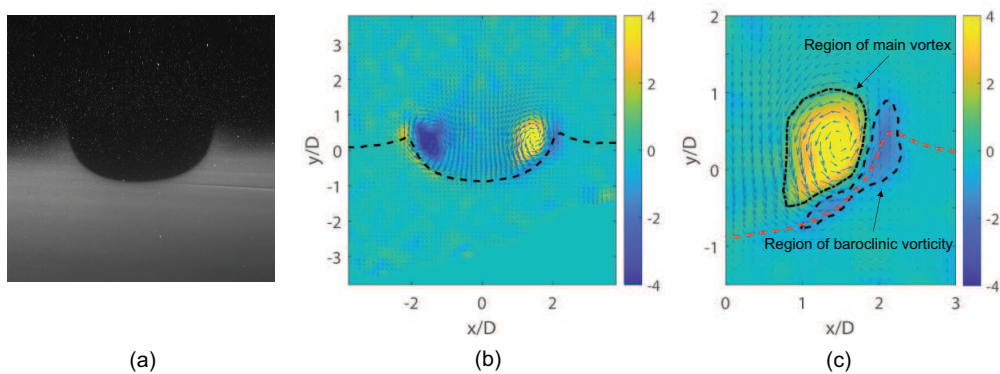}

\caption{(a) Representative PIV image of a vortex ring penetrating the density interface of a stable two-layer system. (b) Velocity field computed by processing the PIV image in panel (a); the color indicates the dimensionless vorticity field, $\boldsymbol{\omega} U_p/D$; the size and direction of the arrows indicate the magnitude and the direction of the velocity field, respectively; the dashed line shows the instantaneous position of the interface between the two fluid layers. (c) shows half of the vortex ring and induced flow in (b). From the vorticity field, the circulation of the main vortex and the baroclinic vorticity are retrieved by numerically integrating within the dotted line region (main vortex) and the black dashed line region (baroclinic vorticity), where the red dashed line indicates the density gradient.}
\label{fig:typical_exp}

\end{figure}

Velocity fields were obtained via two-dimensional Particle Image Velocimetry (PIV)  (Fig.~\ref{fig: setup}(b)). A continuous laser sheet was created using a plano concave cylindrical lens (-3.9 mm focal length) and a laser light beam (1 W, 532 nm, continuous, Laser Glow). Both fluid layers were seeded with 13 $\mu m$ silver-coated hollow glass spheres (Potters Industries Inc). A high-speed camera (Photron Ultima APX-RS) was used at 125 frames per second. The velocity fields were computed using the software Dynamic Studio (Dantec Dynamics) with the single-frame scheme and applying standard validation and filtering algorithms \citep{willert1991}. A representative instantaneous raw PIV image showing a vortex ring penetrating the density interface from the top is presented in Fig.~\ref{fig:typical_exp}(a). Note that the change in contrast is due to the fluorescent dye used to track the interface between the two layers. In this case, the induced velocity and vorticity fields result in the deformation of the interface between the two homogeneous layers (see Fig.~\ref{fig:typical_exp}(b), where the dashed line indicates the position of the density interface defined as the mean pixel value between the top and bottom regions in the raw image).

As the vortex ring crosses the interface, the generation of baroclinic vorticity is evident at the density interface (Fig.~\ref{fig:typical_exp}(b) and (c)) In principle, the production of baroclinic vorticity can be measured experimentally using the density and pressure gradients acquired from the PLIF and the PIV velocity field measurements, respectively. However, the uncertainty associated with this method was too large to ensure accuracy (details can be found in Supplemental Materials). Instead, the circulation of the baroclinic vorticity was calculated by locating the vorticity strand in the vicinity of the sharp density gradient (Fig. \ref{fig:typical_exp}(c)). 
It is important to note that the vortex rings in this study are compact, i.e, the vortex structure has only one separatrix in the cross-sectional area, and there is no additional baroclinic vorticity generation at the inner side of the separatrix (Fig.~\ref{fig:typical_exp}(b) and (c)). According to \cite{norbury1973family} and \cite{palacios2013vortex}, one can classify vortices depending on the value of $\alpha$, which is defined as $\alpha^2 = A / (\pi R_v^2)$, where A is the area of the vortex core and $R_v$ is the vortex mean radius. The value of $\alpha$ ranges from 0 to $\sqrt{2}$, where 0 corresponds to a skinny vortex and $\sqrt{2}$ corresponds to a spherical vortex. In this study, $\alpha$ is approximately 0.55, suggesting that the vortex ring is compact but not close to Hill’s spherical vortex.

\section{Results}\label{sec:results}

\subsection{Vortex ring crossing a density interface}
\label{Sect. discription of the process}

\begin{figure}
\centering
\mbox{\subfigure[Crossing from high to low density]{\includegraphics[width=\textwidth]{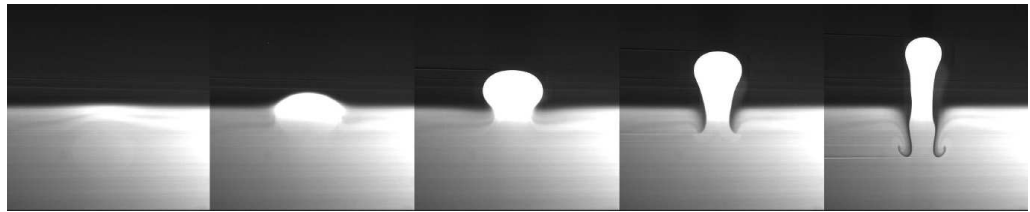}}}\\ 
\mbox{\subfigure[Crossing from low to high density (images flipped upside down)]{\includegraphics[width=\textwidth]{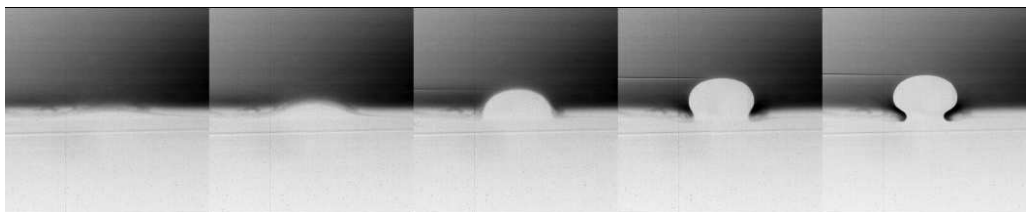}}} \\
\mbox{\subfigure[Position and diameter of vortex rings]{\includegraphics[width=0.45\textwidth]{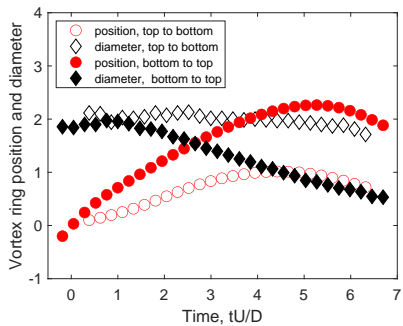}}}
\mbox{\subfigure[Validation]{\includegraphics[width=0.456\textwidth]{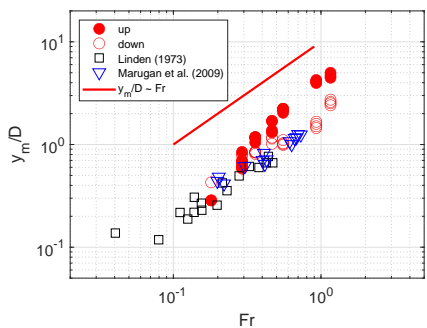}} 
}

\caption{Vortex ring crossing a density interface ($Fr$ = 0.18, $\Delta\rho^*=0.004$). Representative case of a vortex ring moving towards a solution with lower (a) and higher (b) densities. Note that in (b), the images are flipped upside down and the pixel intensities are inverted for clarity. To compare the cases, the corresponding images in (a) and (b) share the same non-dimensional formation time, and $\Delta t U/D=2.5$. (c) Time evolution of the position and diameter of the vortex rings while crossing the interface. (d) Normalized penetration depth as a function of Froude number, $Fr$. Results are compared with relevant studies in literature. The solid line fits the trend of the data in \citet{linden1973} with a slope of 1. } 
\label{fig:experiments}
\end{figure} 

Experiments were conducted using the upper and lower piston-cylinder arrangements in the water tank to produce isolated vortex rings propagating both upwards and downwards or against and along the density gradient (e.g., Fig.~\ref{fig:experiments}(a) and (b), respectively). In both cases, backflow was observed after the vortex ring crossed the interface, in which case some fluid inside the vortex ring was observed to flow back close to its original position (e.g., the last images of Fig.~\ref{fig:experiments}(a) and (b)). This observation is consistent with the experimental and numerical results from \citet{dahm1989} and \citet{Olsthoorn_dalziel_2017}. In this study, the authors observed that whenever a vortex ring crossed a fluid interface, the baroclinic production of vorticity \textit{peeled off} the outer fluid layer of the vortex. Then, due to gravitational effects, this peeled-off layer was transported backward or opposite to the vortex travel direction, thereby creating a backflow once the vortex ring had fully crossed the interface. Also, as the backflow induces an instability of the Kelvin-Helmholtz type, a wavy or rolled-up structure is generated, as seen in Fig.~\ref{fig:experiments}(a) and (b). However, the backflow is much weaker in Fig.~\ref{fig:experiments}(b) than in Fig.~\ref{fig:experiments}(a), presumably due to differences in the direction of travel of each vortex.

In the current study, the position and diameter of the vortex ring were tracked once the vortex ring crossed the interface between the two fluid layers (Fig.~\ref{fig:experiments}(c)). It was observed that the diameter of the vortex ring (solid diamond in Fig.~\ref{fig:experiments}(c))  decreases when crossing the interface to less dense fluids (Fig.~\ref{fig:experiments}(a)). This is likely due to the backflow and the \textit{peeling} process, in line with the experimental work by \citet{dahm1989} and the theoretical prediction in Eqn.~\ref{eqn: vortex ring size}. However, when the vortex ring crosses the interface to denser fluids (hollow diamonds in Fig.~\ref{fig:experiments}(c)), the diameter shrinking effect is smaller than that in the other direction (solid diamonds, bottom to top, Fig.~\ref{fig:experiments}(c)). In addition, the penetration depth of the ring (circles in Fig.~\ref{fig:experiments}(c)) also depends on the direction of travel of the vortex. It was found that the vortex penetrated deeper into the second layer when moving against the density gradient than when moving downward towards the denser fluid solution. The asymmetry in the process can be explained by focusing on the effect of the initial conditions on the generation of baroclinic vorticity as the vortex ring crosses the interface. This will be further discussed in Section 4 using the pressure and density fields.

The same experiment was repeated for different density ratios (Froude number, $Fr$) (Table~\ref{table:props}). The vortex positions were tracked, and the maximum penetration depths were computed and compared with results in the literature (Fig.~\ref{fig:experiments}(d)). Our experimental data (filled and empty red circles) is in good agreement with measurements reported in the literature, thereby validating our methodology. In particular, \citet{linden1973} reported that the slope of the best fit curve was close to 1 (solid line in Fig.~\ref{fig:experiments}(d)).

\subsection{Penetration depth and circulation}
\label{sect: penetration and circulation}

\begin{figure}
\centering
\mbox{\subfigure[Maximum penetration distance]{\includegraphics[width=0.45\textwidth]{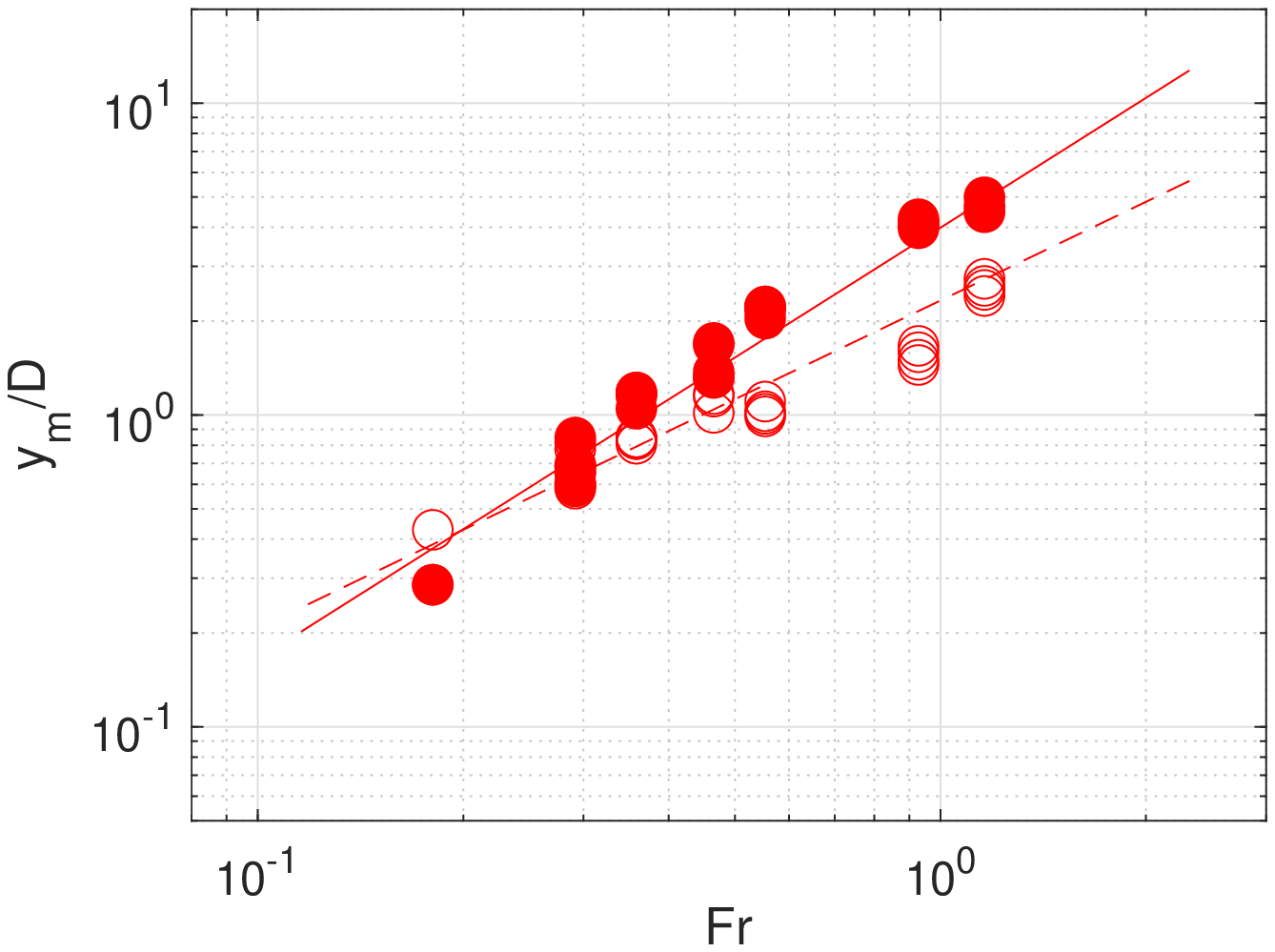}}}
\mbox{\subfigure[Circulation associated with the main vortex ring and with the baroclinic vorticity]{\includegraphics[width=0.456\textwidth]{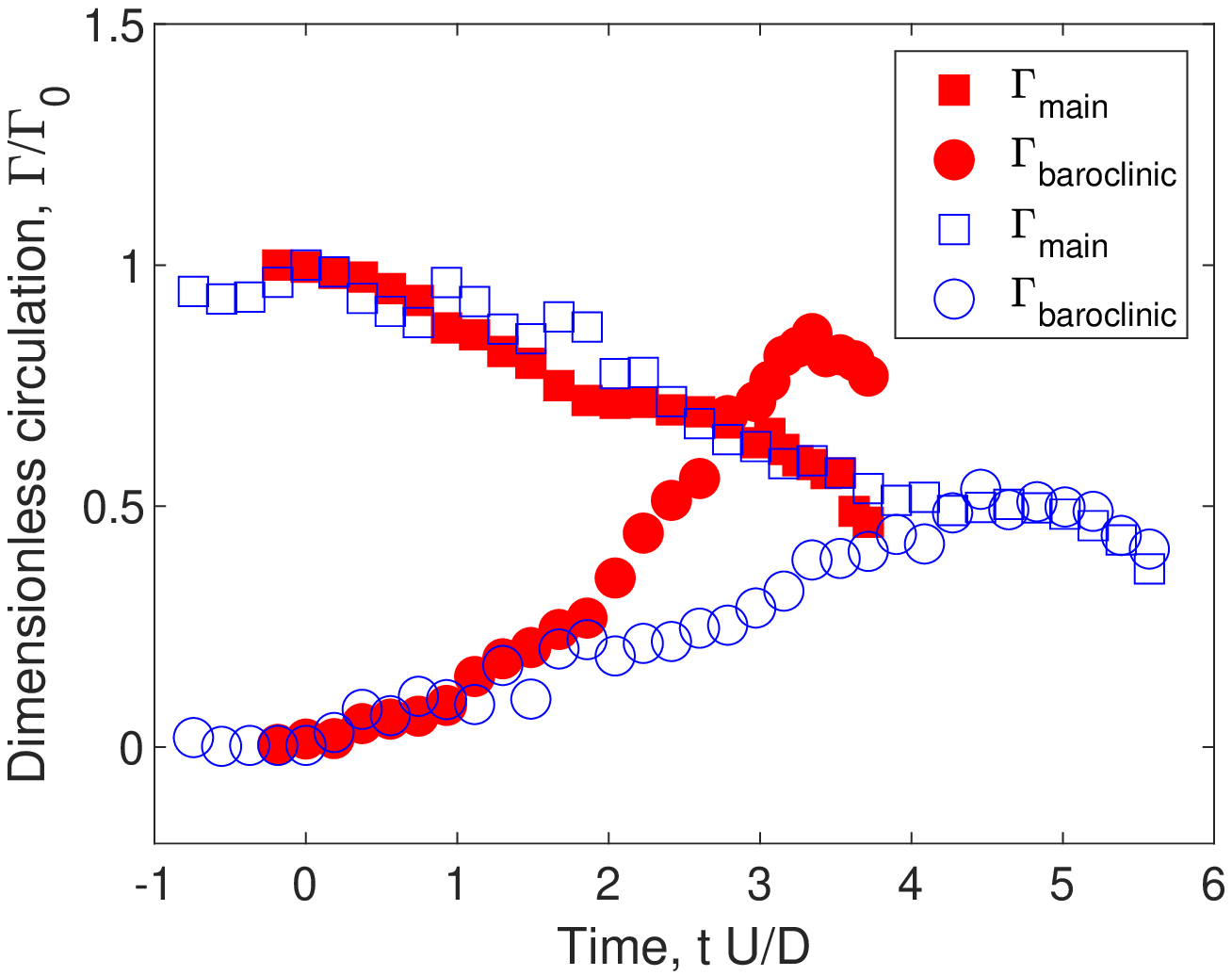}} 
}
\caption{Asymmetry of motion. (a) Normalized penetration distance as a function of Froude number, $Fr$. Empty and filled symbols show experimental results for vortex rings moving along and against the density gradient, respectively. The solid and dashed lines are the best fit for each data group. (b) Normalized circulation, $\Gamma/\Gamma_o$, as a function of dimensionless time for both cases ($\rho = 1005$ kg/m$^3$ in Table~\ref{table:props}). Similar to panel (a), empty and filled symbols correspond to vortex rings moving downwards and upwards, respectively. Squares denote the value of the circulation of the main vortex ring, while the circles correspond to the circulation associated with baroclinic vorticity. } 
\label{fig:penetration_circulation}
\end{figure}

Experiments were repeated with different density ratios (Froude number, $Fr$). The maximum penetration depth is identified and reported for each case (Fig.~\ref{fig:penetration_circulation}(a)). Even though our experimental results follow a general trend as reported in literature, the curves of maximum penetration distance do not overlap. At a given Froude number, the vortex ring moving towards a less dense fluid layer attains a longer penetration depth (solid circles) than if moving towards a denser fluid solution. This is consistent with what has been observed in Sect.~\ref{Sect. discription of the process}, corroborating the inherent asymmetry in the vortex ring motion across a density interface.

To understand the vortex propagation asymmetry, the circulation of the main vortex ring is examined along with the circulation associated with the baroclinic vorticity generated at the density interface (e.g., Fig.~\ref{fig:typical_exp}(b) and (c)). The evolution of circulation for the main vortex ring (squares) and the baroclinic vorticity (circles) once the vortex ring crosses the density interface is presented in Fig.~\ref{fig:penetration_circulation}(b). Independent of the travel direction, the circulation of the main vortex ring decreases when crossing the density interface, which is in line with the \textit{peeling} effects discussed in Sect.~\ref{Sect. discription of the process}. In contrast, the circulation of the baroclinic vorticity (circles) increases when the vortex ring crosses the density interface. It is to be noted that the baroclinic circulation associated with the upward traveling vortex ring (filled circles in Fig.~\ref{fig:penetration_circulation}(b)) reaches its maximum much sooner and attains a much higher value than the downward traveling vortex ring (empty circles in Fig.~\ref{fig:penetration_circulation}(b)). The high baroclinic circulation of the upward-traveling vortex ring is assumed to induce a stronger backflow at the density interface (Fig.~\ref{fig:experiments}(a)), leading to a stronger \textit{peeling} effect and thus resulting in a smaller vortex ring diameter, which is supported by the observations in Fig.~\ref{fig:experiments}(a), (b). Therefore, a smaller drag force is expected for the upward vortex ring, given its smaller diameter, resulting in a greater penetration depth maximum than for the case of a bigger downward-moving vortex ring.


\section{Further analysis and discussion}
\label{further discussion}

\begin{figure}
\centering
\includegraphics[width=0.85\textwidth]{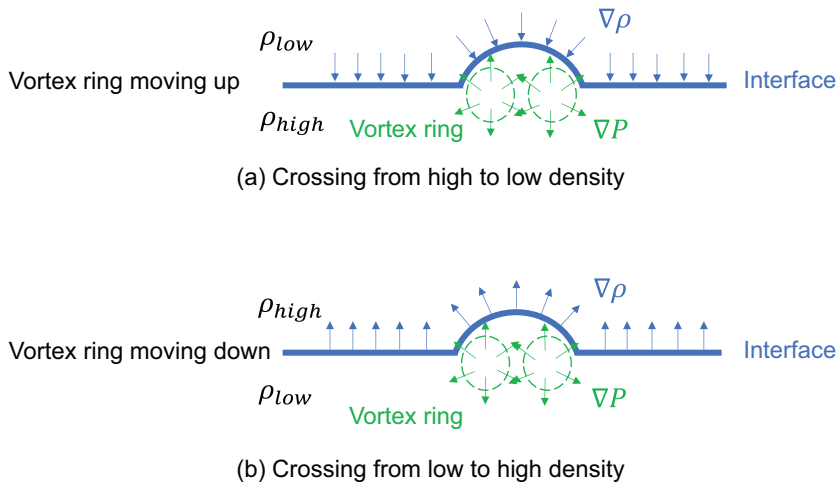}

\caption{Schematic of the physical process causing asymmetry in vortex ring penetration depth. (a) and (b) show the vortex ring crossing from below and above the interface, respectively. The pressure gradient is the same for both cases ($\nabla P_1 = \nabla P_2$). However, the density gradient is different between the two cases ($\nabla\rho_1 = - \nabla\rho_2$). Thus, considering Eqn.~\ref{eqn: baroclinic vorticity} the contribution to baroclinic vorticity production is different: one is positive and the other is negative. } 
\label{fig:intepretation}
\end{figure}

From \citet{marshall2007}, the vorticity conservation equation for a fluid with a non-uniform density is
\begin{equation}
\frac{\partial \boldsymbol{\omega}}{\partial t}+(\boldsymbol{ v} \cdot \nabla) \boldsymbol{\omega}=(\boldsymbol{\omega}\cdot\nabla)\boldsymbol{ v}-\boldsymbol{\omega}(\nabla\cdot\boldsymbol{ v})+\nu\nabla^2\boldsymbol{\omega}+\frac{1}{\rho^2}\left(\nabla\rho\times\nabla P\right),
\end{equation}
where $\boldsymbol{\omega}$ is vorticity, $\boldsymbol{v}$ is velocity, $\nu$ is kinematic viscosity, $\rho$ is density, and $P$ is pressure. For an axisymmetric flow, the first term on the right-hand-side vanishes. For an incompressible fluid flow, the second term on the right hand side is also zero. Therefore, neglecting viscous effects, on either side of the interface gives
\begin{equation}
\frac{D \boldsymbol{\omega}}{D t}=\frac{1}{\rho^2}\left(\nabla\rho\times\nabla P\right). \label{eqn: baroclinic vorticity}
\end{equation}

Considering the axisymmetry of vortex rings, integrating Eqn.~\ref{eqn: baroclinic vorticity} over half the cross-sectional region of the vortex ring and the associated induced flow (as shown in Fig.~\ref{fig:typical_exp}(c)) gives the rate of change of circulation, $\Gamma_b$,
\begin{equation}
\frac{D \Gamma_{b}}{D t}=\int_S\frac{1}{\rho^2}\left(\nabla\rho\times\nabla P\right)d\boldsymbol{S}, \label{eqn: baroclinic circulation}
\end{equation}
where $d\boldsymbol{S}$ is a surface element vector in half the cross-sectional region of the vortex ring and the associated induced flow (as shown in Fig.~\ref{fig:typical_exp}(c)). Eqn.~\ref{eqn: baroclinic circulation} shows that the rate of change of circulation is related to the cross product of the density and pressure gradients at the density interface. 

Fig.~\ref{fig:intepretation} includes a schematic showing the orientation of the density and the pressure gradients as the vortex ring crosses the interface (solid blue line). While the orientation of the pressure gradient is independent of the direction of travel of the vortex ring, the density gradients between the two cases are flipped by $180^{\circ}$, $\Delta\rho^*_{12} = -\Delta\rho^*_{21}$, depending on whether the vortex ring travels towards a denser or lighter fluid layer (Fig.~\ref{fig:intepretation}). Therefore, from Eqn.~\ref{eqn: baroclinic vorticity}, the baroclinic vorticity generated at the density interface has opposite signs, resulting in different contributions to the circulation: one increasing and the other decreasing. Specifically, right to the vortex ring, the cross product of the density and the pressure gradients is positive for the upward penetration (Fig.~\ref{fig:intepretation}(a)) but negative for the downward penetration (Fig.~\ref{fig:intepretation}(b)), suggesting that the total vorticity generated at the density interface is larger for the upward than for the downward penetration case. This trend is consistent with the circulation evolution plot in Fig.~\ref{fig:penetration_circulation}(b).

Finally, it is worth noting that the generation of baroclinic vorticity and the observed difference in the maximum penetration depth of translating vortices in this study are important from an ecological standpoint. While so-called Darwinian drift has been identified as a successful large-scale transport mechanism with potentially far-reaching implications, studies rarely include the effect of migrating direction relative to ocean stratification. Our study suggests that the swimming direction of self-propelled mesozooplankton may have important consequences for the vertical transport of nutrients, carbon, and oxygen across stable density gradients. Taking this into account, from the individual-organism level, the thrust force produced by an organism will likely depend on the swimming direction with respect to the stratification, which may affect organism behaviors in activities such as feeding, preying, and escaping. Likewise, the amount of induced biogenic transport will also depend on the direction in which the organism travels relative to the ocean density gradient. For instance, consider a jellyfish swimming through a stratified fluid, the hydrodynamic forces on the organism and the induced mixing can be expected to differ depending on whether it swims upwards or downwards, even if the jellyfish pumps at the same rate. From the swarm-organism level, those differences inherent from the individual level may become significant during collective motion, such as the diel vertical migrations of aggregations, resulting in different amounts of transport depending on the swimming direction relative to the stratification. Including these differences in either regional or global modeling efforts will be necessary to accurately understand the role of organisms as ecosystem engineers.

\section{Conclusion}
In this study, we performed experiments to quantify the kinematics of a vortex ring crossing the fluid interface of a stable two-layer stratified system. Focus was given to the differences arising from the orientation of the vortex travel direction relative to the density gradient. The maximum penetration depth of the vortex ring crossing a density interface was measured and validated with relevant literature. It was found that the evolution of the vortex ring position, size, and the maximum penetration depth differ with respect to the penetration direction (along or against the density gradient). This difference can be attributed to the asymmetry in the generation of baroclinic vorticity due to the orientation of the density and pressure gradients at the density interface. It was found that a greater maximum penetration depth is associated with a stronger baroclinic vorticity.  Specifically, the higher circulation of the baroclinic vorticity in the upward penetration motion leads to a stronger \textit{peeling} effect on the vortex ring, resulting in a smaller vortex ring size and a smaller drag in the penetration process, which contributes to the larger maximum penetration depth. Finally, while the vortex ring penetration of density interfaces was studied only in the vertical direction, we hypothesize that similar asymmetry can also be observed with the vortex ring penetrating the density interface at an oblique angle or with the vortex ring travelling in a continuously stratified fluid. Going forward, we plan to leverage our findings to improve the characterization of marine ecosystems in global ocean circulation models. While the hydrodynamic signature of migrating mesozooplankton aggregations have just started to get analyzed, symmetry of fluid motion is usually assumed, having important consequences for the net amount of biogenic induced transport of oxygen, carbon, and nutrients. \\

\rule{0.95\textwidth}{1pt}\\
{\bf Acknowledgements.} R.Z. is grateful to the Fulbright-Garca Robles Foundation and PASPA-DGAPA-UNAM for financial support during his sabbatical year at Caltech. M.M.W. thanks John Dabiri for insightful discussions during her years at Caltech. Y.S. and M.M.W. were partially funded by the National Aeronautics and Space Administration Ocean Biology and Biogeochemistry Program (80NSSC22K0284).\\

\noindent
{\bf Funding.} This work was supported by the Fulbright-Garca Robles Foundation and PASPA-DGAPA-UNAM (R.Z.) and the National Aeronautics and Space Administration Ocean Biology and Biogeochemistry Program (Y.S. and M.M.W., grant number 80NSSC22K0284). \\

\noindent
{\bf Supplementary data.} Supplementary materials are available in the supplementary document.\\

\noindent
{\bf Declaration of Interests}. The authors report no conflict of interest.\\

\noindent
{\bf Author ORCIDs.}\\
Y. Su {https://orcid.org/0000-0001-5981-1228} \\
M. M. Wilhelmus {https://orcid.org/0000-0002-3980-2620} \\
R. Zenit {https://orcid.org/0000-0002-2717-4954} \\

\bibliography{Su_etal_2022.bib}
\bibliographystyle{jfm}

\end{document}


\title{Supplemental Materials for Asymmetry of motion: vortex rings crossing a density gradient}

\author
 {Yunxing Su, Monica M. Wilhelmus, Roberto Zenit}

\affiliation
{Center for Fluid Mechanics, Brown University, Providence RI 02912, USA}

\maketitle

\section*{Direct measurement of the baroclinic vorticity}

\begin{figure}
\centering
\mbox{\subfigure[Velocity field by PIV]{\includegraphics[width=0.45\textwidth]{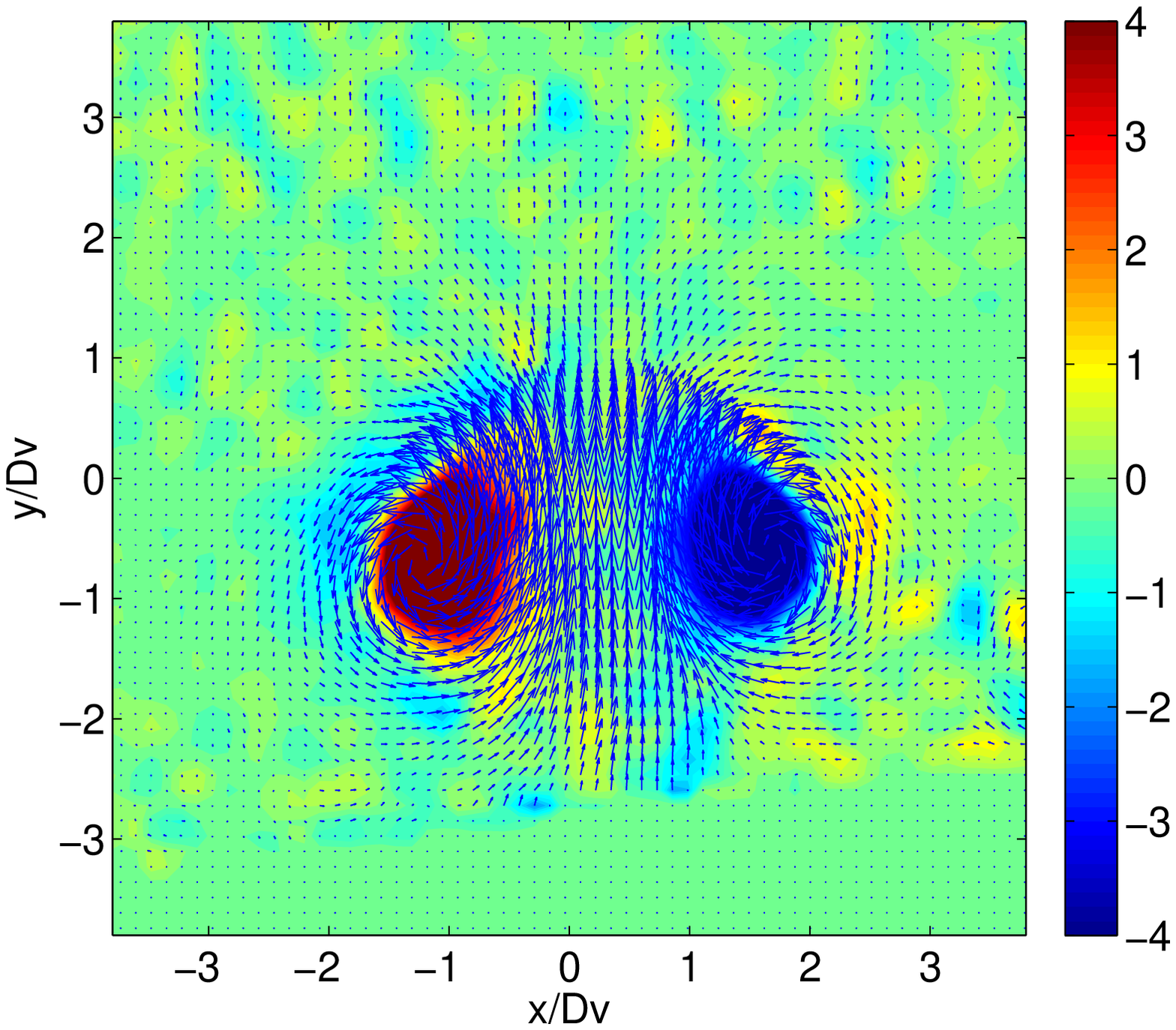}}}
\mbox{\subfigure[Density field by PLIF]{\includegraphics[width=0.45\textwidth]{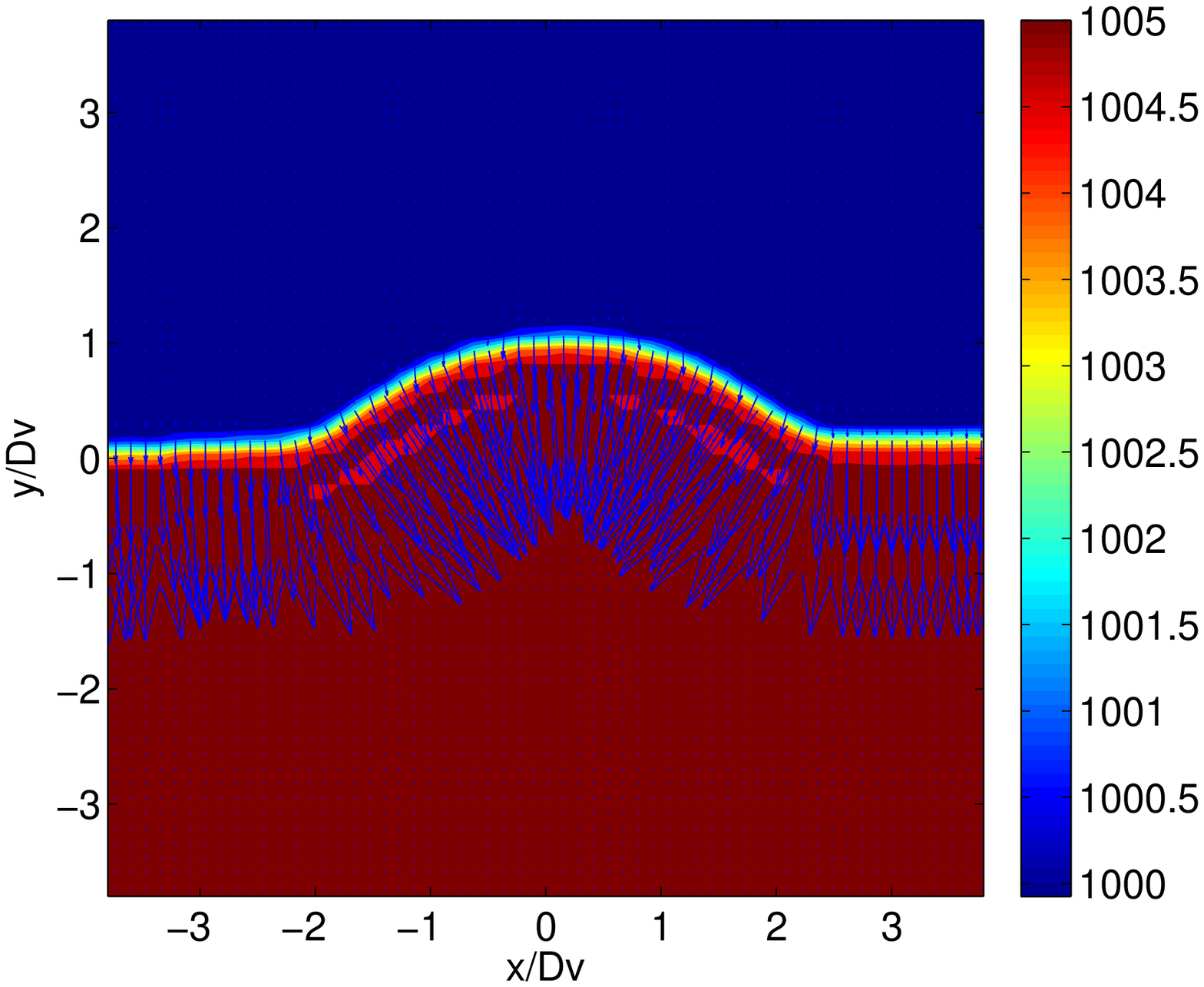}}}
\mbox{\subfigure[Pressure field]{\includegraphics[width=0.45\textwidth]{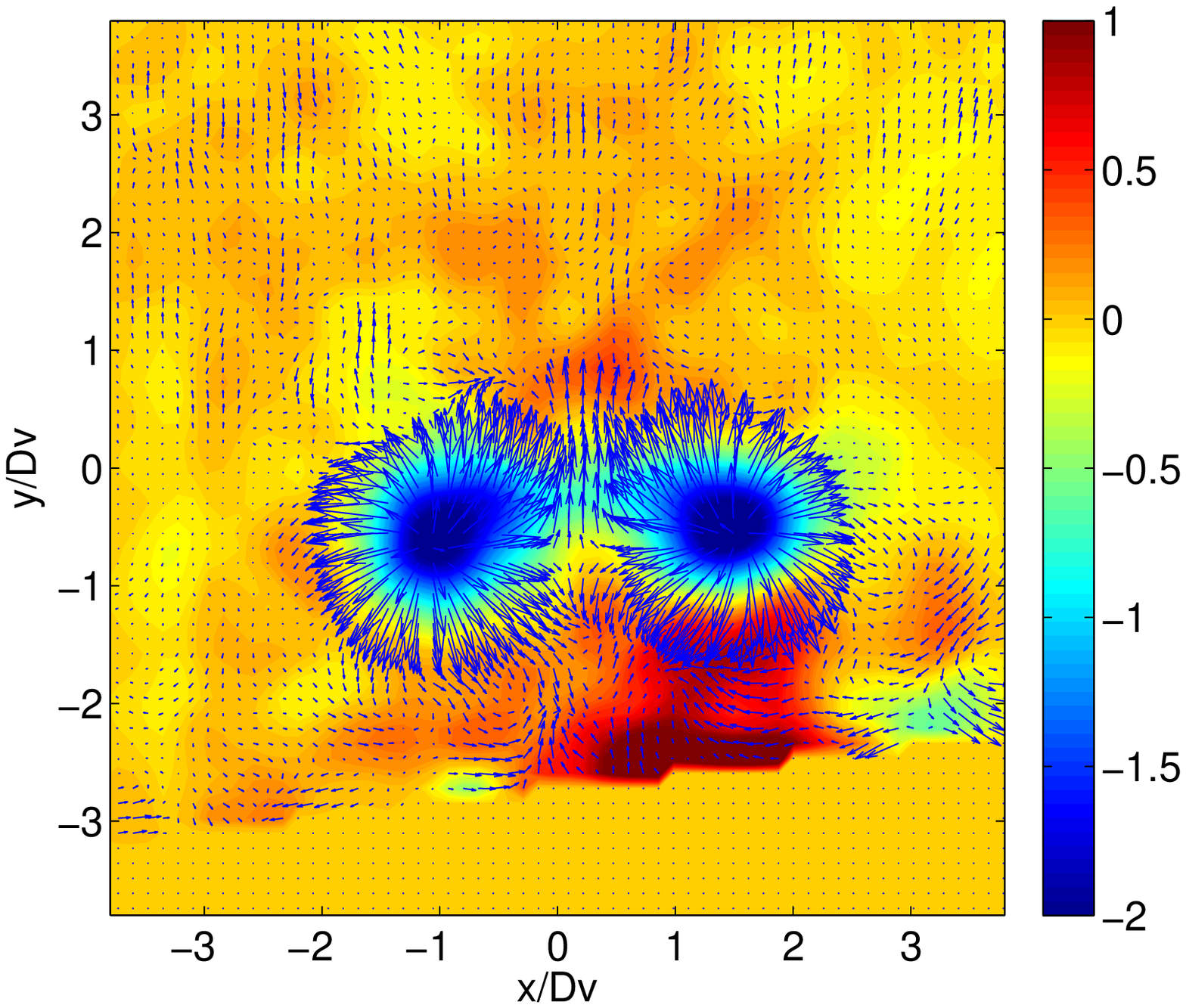}}}
\mbox{\subfigure[Baroclinic vorticity]{\includegraphics[width=0.45\textwidth]{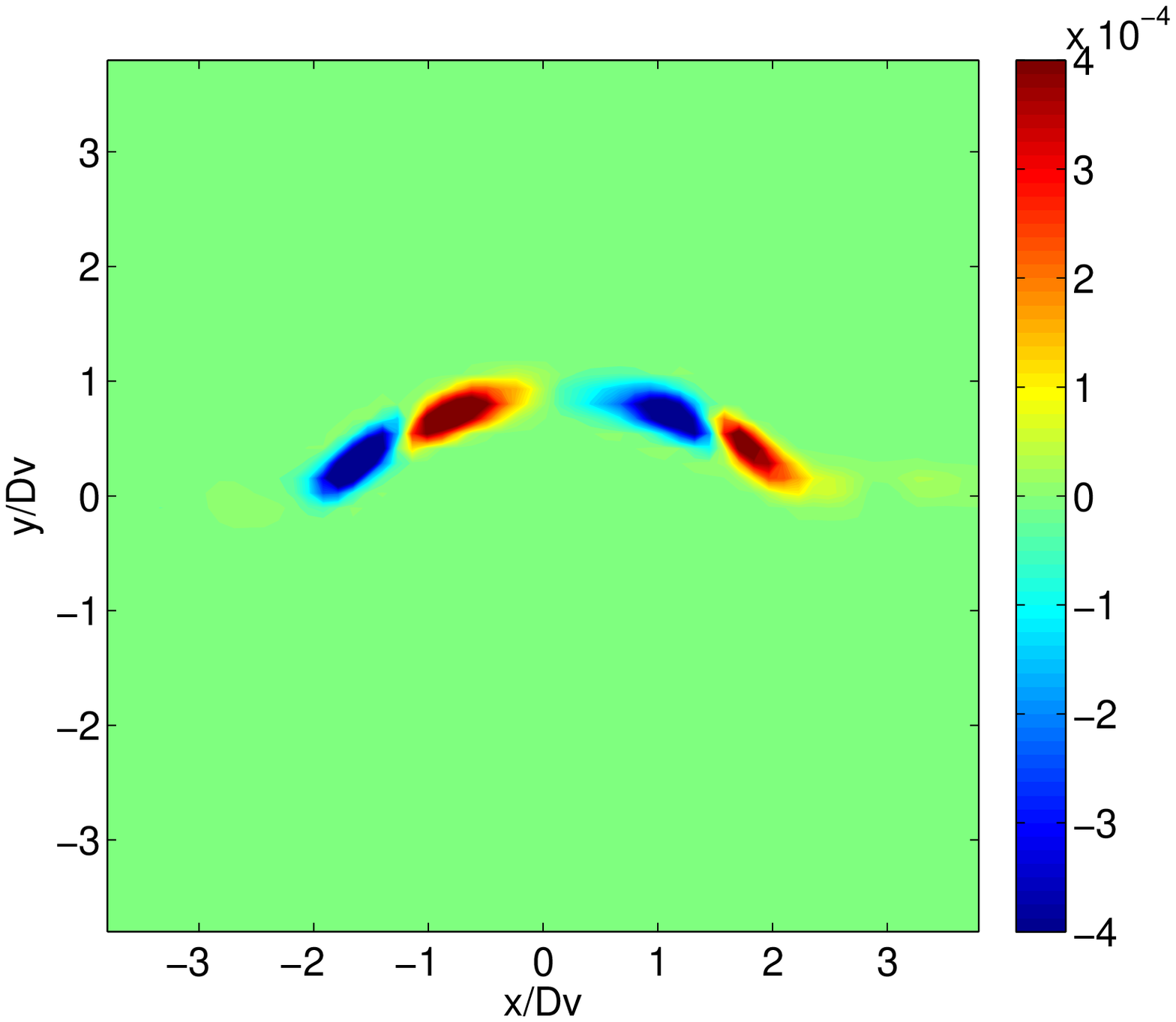}}}
\caption{Typical velocity field (a) and density field (b) of a vortex ring crossing the density interface obtained via PIV and PLIF measurements, respectively. The blue arrows in (b) show the density gradient, and the blue arrows in (c) show the corresponding pressure gradient of the velocity field (a). The associated baroclinic vorticity around the vortex ring is shown in (d), which is calculated from Eqn.~\ref{eqn: baroclinic vorticity} using the velocity and density gradients in (b) and (c). } 
\label{fig:typical_baroclonic}
\end{figure}

In recent years, there have been several studies attempting to conduct simultaneous measurements of both velocity and density fields using Particle Image Velocimetry (PIV) and Planar Laser-Induced Fluorescence (PLIF) (e.g., 
\cite{charogiannis2015}). Such measurements imply the use of two cameras (one for PIV, the other for PLIF) and a series of optical filters and splitters to separate the signals for each measurement. Here, in contrast, simultaneous measurements of both velocity and density fields are obtained from the same imagery of fluids containing small tracer particles as well as fluorescent dye. A representative image is presented in Fig.~\ref{fig:typical_baroclonic}(a).

The density gradient can be readily obtained by tracking the interface using the technique of planar laser-induced fluorescence (PLIF) (as in \cite{crimaldi2008}). To track the evolution of the interface between the two fluids, a small amount (20 ppm) of fluorescent dye (Rhodamine 6G, Sigma-Aldrich) was added to the dense fluid. In this manner, it was possible to track the position of the interface in a plane. Note that in this case, the degree of mixing during the initial instants of the interaction of the vortex ring with the interface is very small. Therefore, the location of the interface was used to calculate the density gradient, $\nabla \rho$, which had a zero value everywhere except at the interface. At the interface, the direction and size of the density gradient vector are given by the shape of the interface and the value of the density contrast. Fig.~\ref{fig:typical_baroclonic}(b) shows a typical measurement of the density around the interface with the arrows showing the density gradients. To calculate the pressure field,  $\nabla P$, the scheme proposed by \cite{dabiri2014} was adopted in which PIV velocity fields are used to solve for the pressure gradient fields from the Navier-Stokes equations. The corresponding pressure field of the velocity field in Fig.~\ref{fig:typical_baroclonic}(a) is shown in Fig.~\ref{fig:typical_baroclonic}(c).

Using the pressure gradient field and density gradient field, according to \citet{marshall2007}, the baroclinic vorticity is calculated as
\begin{equation}
\frac{\partial \omega_B}{\partial t}=\frac{1}{\rho^2}\nabla\rho\times\nabla P \label{eqn: baroclinic vorticity}
\end{equation}
where $\omega_B$ is baroclinic vorticity, $P$ is the pressure, and $\rho$ is the density. The baroclinic vorticity field of the example covered in this section is shown in Fig.~\ref{fig:typical_baroclonic}(d). Note how it is generated at the density interface between the two layers of fluids. 

\bibliography{dense_vortices.bib}
\bibliographystyle{jfm}